\documentclass[graybox]{svmult}

\usepackage{type1cm}        
\usepackage{makeidx}         
\usepackage{graphicx}        
\usepackage{multicol}        
\usepackage[bottom]{footmisc}

\usepackage{newtxtext}       %
\usepackage[varvw]{newtxmath}       

\makeindex             

\usepackage{amsmath}
\usepackage{subcaption}

\usepackage[sort&compress,numbers]{natbib}

\usepackage{tuda-pgfplots}
\usepackage{xcolor}
\pgfplotsset{compat=1.18}

\definecolor{primalDOF}{HTML}{D27FC1}
\definecolor{treeDOF}{HTML}{7FCEC0}

\definecolor{weightOne}{HTML}{7FACD4}
\definecolor{weightTwo}{HTML}{F5B17F}
\definecolor{weightThree}{HTML}{CCE080}
\definecolor{weightFour}{HTML}{B785C1}
\definecolor{weightFive}{HTML}{7F7F7F}

\newcommand{\closure}[1]{\ensuremath{\overline{#1}}}

\newcommand{\dimSpace}[1]{\ensuremath{\operatorname{dim}\left(#1\right)}}

\newcommand{\Curl}{\ensuremath{\operatorname{curl}}}

\newcommand{\trans}{\ensuremath{^{\top}}}

\newcommand{\reluctivity} {\ensuremath{\nu}}
\newcommand{\vecs}[1]{\boldsymbol{#1}}

\newcommand{\vfield}{\ensuremath{\vecs{v}}}
\newcommand{\baseComp}{\ensuremath{\vecs{w}}}
\newcommand{\baseSpace}{\ensuremath{\mathbb{W}}}
\newcommand{\Afield}{\ensuremath{\vecs{A}}}
\newcommand{\Jfield}{\ensuremath{\vecs{J}}}
\newcommand{\Bfield}{\ensuremath{\vecs{B}}}
\newcommand{\nfield}{\ensuremath{\vecs{n}}}

\newcommand{\gDfield}{\ensuremath{\vecs{g}_{\mathrm{D}}}}

\newcommand{\Hcurl}[1]{\ensuremath{{H}(\mathrm{curl};#1)}}
\newcommand{\Hncurl}[1]{\ensuremath{{H}_{0}(\mathrm{curl};#1)}}
\newcommand{\norm}[2]{\ensuremath{\left\vert\left\vert #1 \right\vert\right\vert_{#2}}}
\newcommand{\volPairing}[2]{\ensuremath{\int_{#1}#2\operatorname{dV}}}

\newcommand{\errBfield}{\ensuremath{\epsilon_{\Bfield}}}

\newcommand{\stiffMat}{\ensuremath{\mathbf{K}}}
\newcommand{\couplMat}{\ensuremath{\mathbf{B}}}
\newcommand{\primMat}{\ensuremath{\mathbf{C}_{\mathrm{p}}}}
\newcommand{\coarseMat}{\ensuremath{\mathbf{F}}}
\newcommand{\interfaceMat}{\ensuremath{\mathbf{S}}}
\newcommand{\dofs}{\ensuremath{\mathbf{a}}}
\newcommand{\mults}{\boldsymbol{\lambda}}
\newcommand{\rhs}{\ensuremath{\mathbf{j}}}
\newcommand{\cond}[1]{\ensuremath{\kappa\big(#1\big)}}

\newcommand{\nCoarse}{\ensuremath{n_{\mathrm{gp}}}}

\begin{document}

    \title*{On Domain Decomposition for\\Magnetostatic Problems in 3D}
    \author{Mario Mally\orcidID{0009-0000-2685-3392}, Melina Merkel\orcidID{0000-0002-2104-9167}} 
    \institute{Mario Mally \at Computational Electromagnetics Group and Centre for Computational Engineering, Technische Universität Darmstadt, 64289 Darmstadt, Germany \email{mario.mally@tu-darmstadt.de}
    \and Melina Merkel \at Computational Electromagnetics Group and Centre for Computational Engineering, Technische Universität Darmstadt, 64289 Darmstadt, Germany \email{melina.merkel@tu-darmstadt.de}}
    %
    %
    \maketitle

    \abstract*{The simulation of three dimensional magnetostatic problems plays an important role for example when simulating synchronous electric machines. Building on prior work that developed a domain decomposition algorithm using isogeometric analysis, this paper extends the method to support subdomains composed of multiple patches. This extension enables load-balancing across available CPUs, facilitated by graph partitioning tools such as METIS. The proposed approach enhances scalability and flexibility, making it suitable for large-scale simulations in diverse industrial contexts.}


    \abstract{The simulation of three dimensional magnetostatic problems plays an important role for example when simulating synchronous electric machines. Building on prior work that developed a domain decomposition algorithm using isogeometric analysis, this paper extends the method to support subdomains composed of multiple patches. This extension enables load-balancing across available CPUs, facilitated by graph partitioning tools such as METIS. The proposed approach enhances scalability and flexibility, making it suitable for large-scale simulations in diverse industrial contexts.}

    \section{Introduction}
    \label{sec:intro}
    The design of engineering products often relies on simulations to solve complex problems in both academia and industry. The effectiveness of these workflows relies on the availability of solvers that combine computational efficiency with high accuracy. IsoGeometric Analysis (IGA) has gained popularity as a method capable of meeting these demands. By using Non-Uniform Rational B-Splines (NURBS), IGA allows for precise geometry representation and provides high continuity in the solution space. This makes IGA particularly suitable for magnetostatic simulations, e.g., for the simulation of electric machines, where the geometry is cylindrical, and quantities of interest like the mechanical torque have to be computed with high accuracy. Research into spline spaces for electromagnetic field representation, e.g., \cite{Buffa_2010aa,Buffa_2015aa,Buffa_2019ac}, has allowed the use of IGA for such applications as in \cite{Merkel_2022ab}.

    As computational models grow in size and complexity, parallelized approaches become increasingly important. Domain Decomposition (DD) methods help by splitting the simulation domain into smaller, manageable subdomains and distributing the computational work accordingly. Tearing and Interconnecting (TI) strategies, such as the well-established Finite Element Tearing and Interconnecting (FETI) and the Isogeometric Tearing and Interconnecting (IETI) approaches, are particularly promising for their scalability in large-scale problems. These methods employ Lagrange multipliers to couple subdomains and promise scalability for large problems \cite{Farhat_1991aa,Kleiss_2012ab,Bouclier_2022aa}. To exploit parallelism, the subdomain-related problems shall be individually solvable, particularly those without sufficient boundary conditions on floating subdomains. This challange is typically tackled by a dual-primal approach \cite{Farhat_2001aa,Klawonn_2002aa} or a total/all-floating approach \cite{Dostal_2006aa,Of_2009ab,Bosy_2020aa}.

    Magnetostatic simulations, mostly modeled using scalar potential formulations, have been approached using TI-based methods before \cite{Marcsa_2013aa,Ghenai_2024aa,Schneckenleitner_2022aa}. The vector potential formulation presents additional difficulties due to the non-uniqueness introduced by the kernel of the $\Curl$-operator. Efforts to address this, such as tree-cotree decompositions and regularization techniques, have been proposed in \cite{Toselli_2006aa,Yao_2012aa,Albanese_1988aa,Kapidani_2022aa}.

    This paper builds on and extends \cite{Mally_2024ae} by introducing an automated domain decomposition framework using the graph-partitioning software METIS \cite{Karypis_1997aa}. Unlike traditional IETI approaches, which directly work on a patch-based decomposition, our method allows multipatch regions as subdomains, enabling load-balancing for large-scale simulations. The explicit tree-cotree algorithm of \cite{Mally_2024ae} is analyzed and extended to the new subdomain configuration. Using the \texttt{GeoPDEs} library \cite{Vazquez_2016aa}, we demonstrate the performance of our approach through numerical experiments, showcasing its scalability in solving magnetostatic problems.

    The paper is structured as follows. In Sec.~\ref{sec:fundamentals}, we introduce all necessary fundamentals to understand and extend the dual-primal IETI approach from \cite{Mally_2024ae} to the more general setting of this paper. The first part is concerned about the decomposition and discretization. Starting in Sec.~\ref{sec:dualPrimal} an overview of the dual-primal approach is given. Sec.~\ref{sec:treeCotree} goes into detail how tree DOFs and primal DOFs need to be selected to obtain local invertibility. At last, the analytical findings are verified in the numerical experiments in Sec.~\ref{sec:numExp}.

    \section{Dual-Primal Tearing and Interconnecting for Magnetostatics}\label{sec:fundamentals}
    We begin by introducing the vector potential approach for magnetostatic problems
    \begin{equation}
        \Curl\left(\reluctivity\Curl\Afield\right)=\Jfield,\label{eq:strongMS}
    \end{equation}
    with the magnetic vector potential \Afield, the excitation current density \Jfield and the magnetic reluctivity \reluctivity, on a bounded, simply-connected domain $\Omega\subset\mathbb{R}^3$ where the magnetic flux density is represented by $\Bfield=\Curl\Afield$ \cite{Jackson_1998aa}. For brevity, we only impose Dirichlet conditions, i.e., $\Afield\times\nfield = \gDfield$, on the boundary $\partial\Omega$, although mixed Dirichlet-Neumann configurations are possible \cite{Mally_2024ae}. Without loss of generality, we focus on $\gDfield=\boldsymbol{0}$ corresponding to perfect electric conducting boundary conditions. In this formulation, it is necessary to handle the nontrivial kernel of the $\Curl$-operator consisting of gradient fields. To mitigate the influence of the kernel, the problem is typically gauged by explicitly or implicitly prescribing the kernel components. The remaining components, orthogonal to the kernel, are then computed from \eqref{eq:strongMS}. This idea can be incorporated both in the continuous setting, e.g, with a typical Coulomb gauge, as well as in the discrete setting, e.g, with tree-cotree gauging. We focus on the latter and in the following sections describe the discretization and domain decomposition before detailing an appropriate tree-cotree gauge.

    The weak formulation corresponding to \eqref{eq:strongMS} consists of finding $\Afield\in\Hncurl{\Omega}$ such that
    \begin{equation}
        \volPairing{\Omega}{(\reluctivity\Curl\Afield)\cdot(\Curl\vfield)}=\volPairing{\Omega}{\Jfield\cdot\vfield}\label{eq:weakMS}
    \end{equation}
    for all $\vfield\in\Hncurl{\Omega}$, where $\Hncurl{\Omega}$ denotes the adequate Hilbert space for vector potentials with vanishing tangential component on the boundary \cite{Monk_2003aa}. Discretizing using a Galerkin approach, we approximate $\Hncurl{\Omega}$ with a finite-dimensional subspace $\baseSpace = \operatorname{span}(\baseComp_i)_{i=1}^{n_{\mathrm{g}}}\subset\Hncurl{\Omega}$, and represent the solution as a linear combination of basis functions $\Afield=\sum_{i=1}^{n_{\mathrm{g}}} a_i\baseComp_i$. This yields the discrete system
    \begin{equation}
        \stiffMat_{\mathrm{g}}\dofs_{\mathrm{g}} = \rhs_{\mathrm{g}}\label{eq:discrMS}
    \end{equation}
    with
    \begin{align*}
        \left(\stiffMat_{\mathrm{g}}\right)_{ij} &= \volPairing{\Omega}{\reluctivity\left(\Curl\baseComp_j\right)\cdot\left(\Curl\baseComp_i\right)}\\
        \left(\rhs_{\mathrm{g}}\right)_i &= \volPairing{\Omega}{\Jfield\cdot\baseComp_i}
    \end{align*}
    which represents the discretization of problem \eqref{eq:weakMS}. For the  discrete space $\baseSpace$, we employ the same $\Curl$-conforming IGA multipatch space as in \cite{Mally_2024ae}. There the domain is decomposed into multiple nonoverlapping patches $\Omega_{\mathrm{mp}}^{(j)}$ such that $\closure{\Omega}=\bigcup_{j=1}^{N_{\mathrm{mp}}}\closure{\Omega_{\mathrm{mp}}^{(j)}}$ and $\Omega_{\mathrm{mp}}^{(j)}\cap\Omega_{\mathrm{mp}}^{(k)}=\emptyset$ for all $j\neq k$ and $j,k\in\{1,\ldots,N_{\mathrm{mp}}\}$. We assume that every patch can be represented through the regular mapping $\boldsymbol{F}^{(j)}\colon\hat{\Omega}\rightarrow\Omega_{\mathrm{mp}}^{(j)}$ with reference domain $\hat{\Omega}=\left(0,1\right)^3$. In the following, we call the six (open) boundary sides of $\hat{\Omega}$, that are mapped via $\boldsymbol{F}^{(j)}$, patch-facets. We further define a patch-interface as $\Gamma_{\mathrm{mp}}^{(jk)}=\partial\Omega_{\mathrm{mp}}^{(j)}\cap\partial\Omega_{\mathrm{mp}}^{(k)}$ and restrict them to only consist of two fully shared patch-facets from $\Omega_{\mathrm{mp}}^{(j)}$ and $\Omega_{\mathrm{mp}}^{(k)}$.

\begin{figure}
        \centering
        \begin{subfigure}[b]{0.49\linewidth}
            \centering
            \includegraphics[width=0.85\linewidth,height=0.82\linewidth]{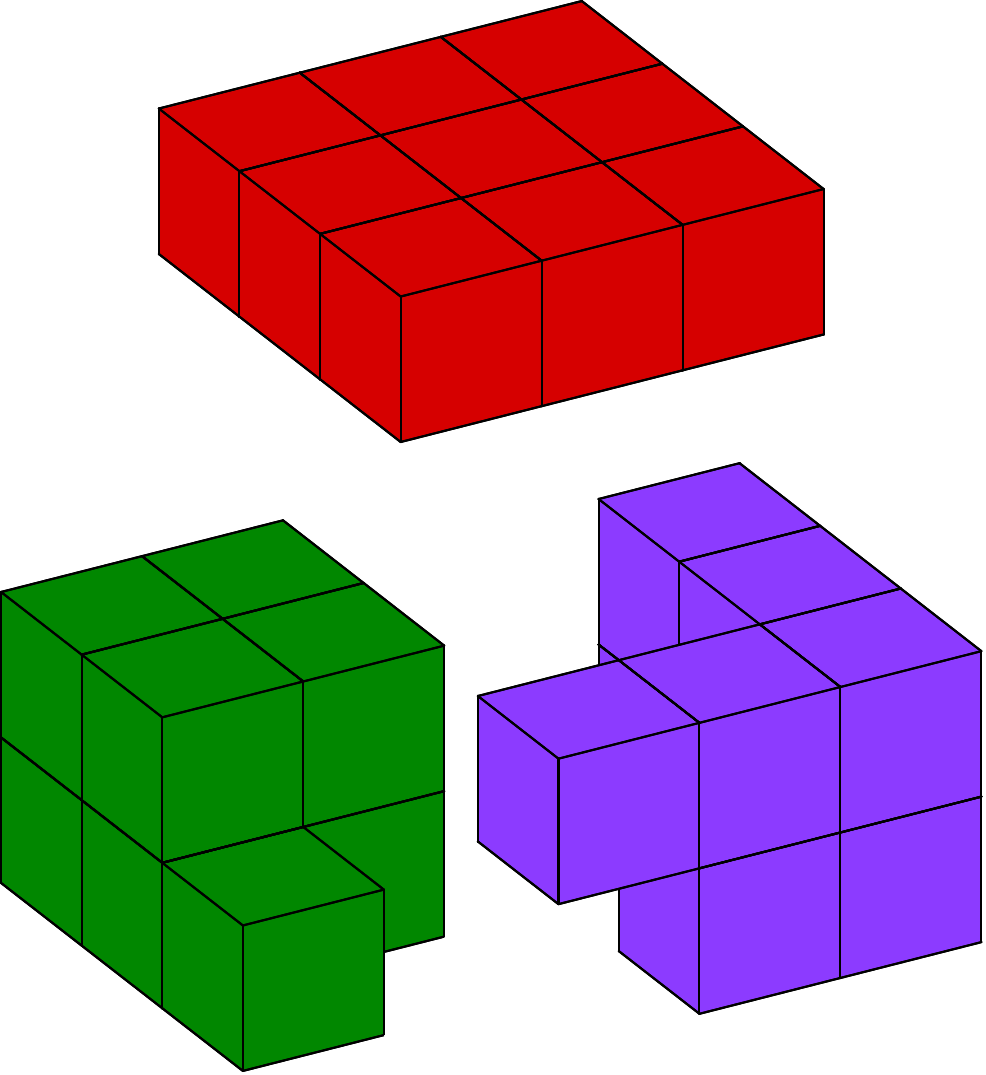}
            \caption{Colored subdomains.}
            \label{fig:decomp_a}
        \end{subfigure}
        \begin{subfigure}[b]{0.49\linewidth}
            \centering
            \includegraphics[width=0.9\linewidth]{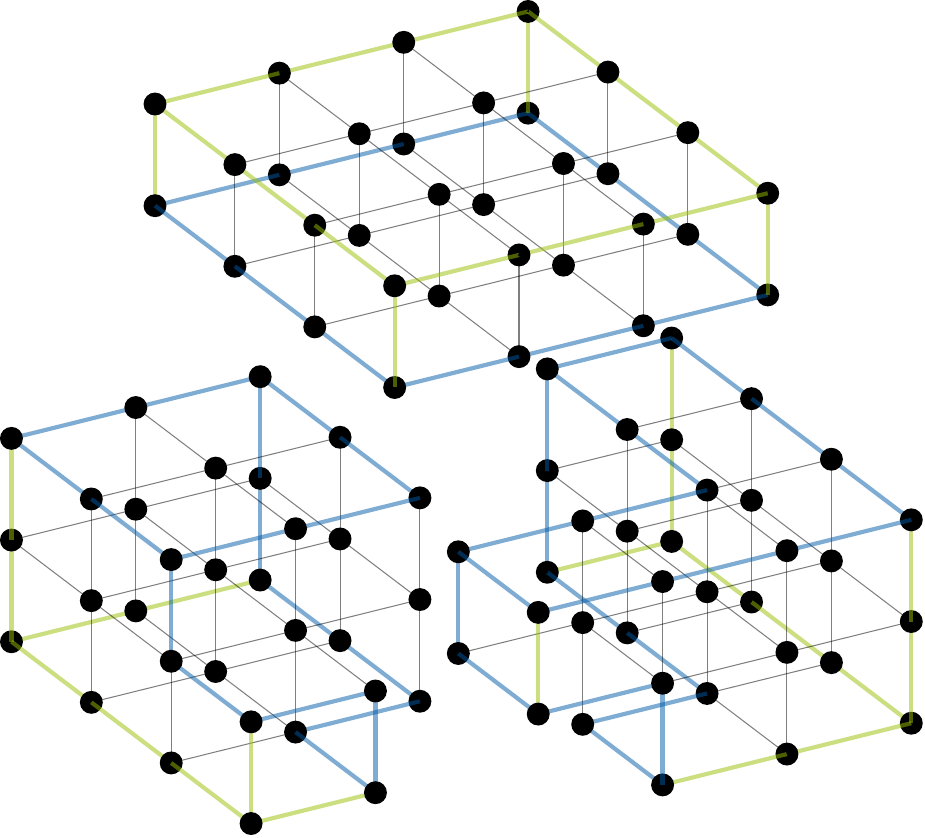}
            \caption{Colored edges of wire basket.}
            \label{fig:decomp_b}
        \end{subfigure}
        \caption{Exemplary decomposition (exploded view) of a cube $\Omega=\left(0,3\right)^3$ consisting of 27 patches, which includes a visualization of the wire basket. The edges in the wire basket are marked in blue and green. Blue refers to those originating from interfaces while green denotes the parts of the wire basket originating from splitting the remaining subdomain boundaries into multiple facets.}
        \label{fig:decomp}
    \end{figure}

    The $N_{\mathrm{mp}}$ patches are grouped into $N_{\mathrm{sub}}\leq N_{\mathrm{mp}}$ connected subdomains $\Omega^{(k)}$ which satisfy $\closure{\Omega^{(k)}}=\bigcup_{j\in\mathcal{I}_{k}}\closure{\Omega_{\mathrm{mp}}^{(j)}}$. Here, $\mathcal{I}_{k}$ denotes the index set of patches belonging to subdomain $\Omega^{(k)}$. An exemplary decomposition is shown in Fig.~\ref{fig:decomp}. We additionally restrict ourselves to configurations, where each subdomain is simply-connected to avoid dealing with additional kernel elements \cite{Arnold_2018aa}. In our implementation, subdomain configurations are generated using METIS, which partitions the dual graph of the multipatch structure. In this graph, each patch corresponds to a dual-node and each patch-interface to a dual-edge. However, note that while METIS can guarantee connectedness, it is not able to enforce simple-connectedness.

    For interfaces between subdomains, we define $\Gamma^{(kl)}=\partial\Omega^{(k)}\cap\partial\Omega^{(l)}$, which corresponds to a union of patch-interfaces or correspondingly a union of patch-facets. For the following algorithms, we divide $\partial\Omega^{(k)}$ into smaller parts, that we call facets. We assume that these facets are unions of patch-facets and that every interface is considered to be a facet. For the remaining Dirichlet boundary, each connected component can be considered a facet, and it may be beneficial to further subdivide the Dirichlet facets, e.g., to prescribe different boundary conditions. In our implementation, we enforce that two patch-facets of the same patch sharing a boundary segment belong to two separate facets.

    Finally, we define the term wire basket to refer to all edges and nodes in the (control) mesh, whose basis functions do not vanish on the union of all facet boundaries. In other words, the wire basket consists of ``cross-edges'' (correspondingly ``cross-points'') shared by more than two subdomains, edges/nodes located at the intersection of interfaces and the Dirichlet boundary as well as the remaining intersections of facet boundaries. Fig.~\ref{fig:decomp_b} provides a visualization for an appropriate wire basket of the configuration.

    Using the weak formulation \eqref{eq:weakMS}-\eqref{eq:discrMS}, we construct the block-diagonal stiffness matrix $\stiffMat=\operatorname{diag}\left(\stiffMat^{(1)},\ldots,\stiffMat^{(N_{\mathrm{sub}})}\right)\in\mathbb{R}^{n\times n}$ with subdomain component $\stiffMat^{(k)}\in\mathbb{R}^{n_k\times n_k}$ of $\Omega^{(k)}$. Incorporating the classical TI-constraints $\couplMat\dofs=\mathbf{0}$, where $\couplMat\in\{-1,0,1\}^{m\times n}$ is a signed boolean matrix, we arrive at
    \begin{equation}
        \begin{bmatrix}
            \stiffMat & \couplMat\trans \\
            \couplMat & \mathbf{0} \\
        \end{bmatrix}\begin{bmatrix}
            \dofs \\
            \mults \\
        \end{bmatrix}=\begin{bmatrix}
            \rhs \\
            \mathbf{0} \\
        \end{bmatrix}\label{eq:fetiNaive}
    \end{equation}
    with Lagrange multipliers $\mults\in\mathbb{R}^{m}$. To resolve ill-posedness of \eqref{eq:fetiNaive} due to discrete gradient fields and problematic coupling constraints for DOFs shared by more than two subdomains, we employ a dual-primal approach combined with tree-cotree gauging.

    \subsection{The Dual-Primal Approach}\label{sec:dualPrimal}
    In this section, we describe the dual-primal approach from a general perspective to provide context and analyze which essential properties are not yet satisfied and need to be enabled with the help of a tree-cotree gauge. In dual-primal TI, one splits the DOFs into primal ones $\dofs_{\mathrm{p}}\in\mathbb{R}^{n_{\mathrm{p}}}$, for which continuity is not enforced via constraints but explicitly incorporated in the system, and the remaining DOFs $\dofs_{\mathrm{r}}\in\mathbb{R}^{n_{\mathrm{r}}}$. The explicit continuity can be expressed by substituting $\dofs_{\mathrm{p}}=\primMat\mathbf{p}$, where  $\couplMat_{\mathrm{p}}\primMat=\mathbf{0}$, $\primMat\in\{0,1\}^{n_{\mathrm{p}} \times \nCoarse}$, and  $n_{\mathrm{gp}}=\dimSpace{\ker\left(\couplMat_{\mathrm{p}}\right)}$. Here, the coupling matrix is partitioned as $\couplMat=\begin{bmatrix} \couplMat_{\mathrm{r}} & \couplMat_{\mathrm{p}} \end{bmatrix}$. Substituting this into the naive TI formulation \eqref{eq:fetiNaive} leads to the modified global system
    \begin{equation}
        \begin{bmatrix}
            \stiffMat_{\mathrm{rr}} & \stiffMat_{\mathrm{rp}}\primMat & \couplMat_{\mathrm{r}}\trans  \\
            \primMat\trans\stiffMat_{\mathrm{pr}} & \primMat\trans\stiffMat_{\mathrm{pp}}\primMat & \mathbf{0} \\
            \couplMat_{\mathrm{r}} & \mathbf{0} & \mathbf{0} \\
        \end{bmatrix}\begin{bmatrix}
            \dofs_{\mathrm{r}} \\
            \mathbf{p} \\
            \mults_{\mathrm{r}} \\
        \end{bmatrix}=\begin{bmatrix}
            \rhs_{\mathrm{r}} \\
            \primMat\trans\rhs_{\mathrm{p}} \\
            \mathbf{0} \\
        \end{bmatrix}\label{eq:gloDP}
    \end{equation}
    where redundant constraints were additionally removed such that $\mults_{\mathrm{r}}\in\mathbb{R}^{m_{\mathrm{r}}}$ are the remaining multipliers. These steps effectively reduce each set of coupled DOFs in $\dofs_{\mathrm{p}}$ to one single DOF in $\mathbf{p}$.

    The stiffness matrix of remaining DOFs is block-diagonal, i.e.,
    \begin{equation*}
        \stiffMat_{\mathrm{rr}}=\operatorname{diag}\left(\stiffMat_{\mathrm{rr}}^{(1)},~\ldots~,~\stiffMat_{\mathrm{rr}}^{(N_{\mathrm{sub}})}\right)
    \end{equation*}
    and it is typical to assume that every $\stiffMat_{\mathrm{rr}}^{(k)}$ is invertible. However in our setting, all $\stiffMat_{\mathrm{rr}}^{(k)}$ still have a non-trivial kernel due to local gradient fields. The combination of the tree-cotree gauge and the selection of primal DOFs must ensure that the influence of local gradient fields is fully eliminated. Once $\stiffMat_{\mathrm{rr}}$ is invertible, $\dofs_{\mathrm{r}}$ can be eliminated from \eqref{eq:gloDP} with
    \begin{equation}
        \dofs_{\mathrm{r}} = \stiffMat_{\mathrm{rr}}^{-1}\left(\rhs_{\mathrm{r}} - \stiffMat_{\mathrm{rp}}\primMat\mathbf{p} - \couplMat_{\mathrm{r}}\trans\mults_{\mathrm{r}} \right).\label{eq:recRem}
    \end{equation}
    Exploiting this in the dual-primal TI context leads to
    \begin{align}
        \interfaceMat\mults_{\mathrm{r}} = \left(\mathbf{G}\coarseMat^{-1}\mathbf{G}\trans-\mathbf{W}\right)\mults_{\mathrm{r}} &= \mathbf{G}\coarseMat^{-1}\mathbf{d} - \mathbf{e}\label{eq:intProb} \\
        \mathbf{a}_{\mathrm{p}} &= \primMat\coarseMat^{-1}\left(\mathbf{d} - \mathbf{G}\trans\mults_{\mathrm{r}}\right)\label{eq:coarseProb} \\
        \dofs_{\mathrm{r}} &= \stiffMat_{\mathrm{rr}}^{-1}\left(\rhs_{\mathrm{r}} - \stiffMat_{\mathrm{rp}}\mathbf{a}_{\mathrm{p}} - \couplMat_{\mathrm{r}}\trans\mults_{\mathrm{r}} \right).\label{eq:localProb}
    \end{align}
    as the final procedure to compute a discrete solution, following \cite{Farhat_2001aa}. In this setting,
    \begin{align*}
        \coarseMat&=\primMat\trans\stiffMat_{\mathrm{pr}}\stiffMat_{\mathrm{rr}}^{-1}\stiffMat_{\mathrm{rp}}\primMat-\primMat\trans\stiffMat_{\mathrm{pp}}\primMat, \\
        \mathbf{G}&=\couplMat_{\mathrm{r}}\stiffMat_{\mathrm{rr}}^{-1}\stiffMat_{\mathrm{rp}}\primMat, \\
        \mathbf{W}&=\couplMat_{\mathrm{r}}\stiffMat_{\mathrm{rr}}^{-1}\couplMat_{\mathrm{r}}\trans, \\
        \mathbf{d}&=\primMat\trans\stiffMat_{\mathrm{pr}}\stiffMat_{\mathrm{rr}}^{-1}\rhs_{\mathrm{r}} - \primMat\trans\rhs_{\mathrm{p}},\\
        \mathbf{e}&=\couplMat_{\mathrm{r}}\stiffMat_{\mathrm{rr}}^{-1}\rhs_{\mathrm{r}}.
    \end{align*}
    are the corresponding matrices and vectors used in \eqref{eq:intProb}-\eqref{eq:localProb}. The possibility of employing parallel computations mainly originates from the block-structure of $\stiffMat_{\mathrm{rr}}$ which can be exploited when computing the action of $\stiffMat_{\mathrm{rr}}^{-1}$. For further details, the reader is referred to \cite{Farhat_2001aa}.

    Typically, the interface problem \eqref{eq:intProb} is solved using iterative methods such as the preconditioned conjugate gradient (PCG) method, e.g. see \cite{Farhat_2000aa,Farhat_2001aa}. To increase scalability an appropriately designed preconditioner is required. Following the approach of \citeauthor{Farhat_2000aa} in \cite{Farhat_2000aa}, we investigate the Dirichlet preconditioner
    \begin{equation}
        \mathbf{M}_{\mathrm{D}}^{-1}=\mathbf{D}_{\mathrm{s}}\couplMat_{\mathrm{r_{I}}}\interfaceMat_{\mathrm{r_{I}}\mathrm{r_{I}}}\couplMat_{\mathrm{r_{I}}}\trans\mathbf{D}_{\mathrm{s}}\quad\text{with}\quad\interfaceMat_{\mathrm{r_{I}}\mathrm{r_{I}}}=\stiffMat_{\mathrm{r_{I}}\mathrm{r_{I}}} - \stiffMat_{\mathrm{r_{I}}\mathrm{r_{V}}}\stiffMat_{\mathrm{r_{V}}\mathrm{r_{V}}}^{-1}\stiffMat_{\mathrm{r_{V}}\mathrm{r_{I}}}.\label{eq:DirPrec}
    \end{equation}
    In \eqref{eq:DirPrec}, the remaining DOFs $\dofs_{\mathrm{r}}$ are split into $\dofs_{\mathrm{r_{I}}}$ and $\dofs_{\mathrm{r_{V}}}$, where $\dofs_{\mathrm{r_{I}}}$ are the DOFs associated with edges on an interface and $\dofs_{\mathrm{r_{V}}}$ are unconstrained DOFs not associated with interfaces. Here, $\mathbf{D}_{\mathrm{s}}$ is a diagonal scaling matrix used to account for material discontinuities between subdomains, \cite{Farhat_2000aa,Klawonn_2002aa}. In our numerical examples, we focus on problems with a homogeneous material distribution. Consequently, we assume $\mathbf{D}_{\mathrm{s}}=\mathbf{I}$ and omit further discussion on the diagonal scaling.

    \subsection{Tree-Cotree Gauging and Selection of Primal DOFs}\label{sec:treeCotree}
    Tree-cotree gauging aims to determine which DOFs in a discrete problem are fixed by the singular linear equation system and which remain undetermined. This information enables the construction of generalized inverses which specify a solution up to the kernel contribution. This contribution is either prescribed implicitly or explicitly using a gauge condition. In the context of 3D $\Curl$-$\Curl$ problems, the undetermined components can be identified by constructing a spanning tree on the mesh, provided that the DOFs correspond to edges of the mesh. The tree edges then correspond to the undetermined DOFs and the remaining cotree edges correspond to the determined ones \cite{Albanese_1988aa,Manges_1995aa}.

    To incorporate this idea into the TI method and ensure both local and global solvability, we construct a special spanning tree at the global level, i.e., \eqref{eq:discrMS} as in \cite{Mally_2024ae}. The constructed spanning tree is then projected onto the subdomain-graphs. This preserves the consistency of the coupling constraints because the tree edges match across subdomain interfaces by construction. Additionally, the spanning tree suffices to correctly gauge the corresponding multipatch problem associated with \eqref{eq:discrMS}, as shown in \cite{Kapidani_2022aa}.

    Another important aspect is the consistency with respect to the Dirichlet boundary. When eliminating DOFs, the kernel size of the reduced matrix decreases accordingly. In our setting, the number of kernel elements reduces by the amount of tree edges of a tree that was only grown on the Dirichlet boundary mesh. Therefore, the global tree must also remain a tree when restricted to the Dirichlet boundary mesh \cite{Dular_1995aa}.

    To obtain invertible submatrices $\stiffMat_{\mathrm{rr}}^{(k)}$, we impose additional requirements on the spanning tree. Following \cite{Mally_2024ae} for the problem setting of this paper, the tree construction proceeds as follows: First, the tree is grown on the wire basket. Next, the tree is extended into the local facets and finally, it is extended into the interiors of all subdomains.

    The consistency with respect to the Dirichlet boundary can be obtained by additionally splitting the hierarchy on the wire basket components. First, the tree is grown on the intersection of interfaces and Dirichlet boundary. This effectively separates interfaces and Dirichlet boundary and allows treating both facet types independently of each other in the tree construction. Consequently, consistency with the Dirichlet BCs can be ensured because the constructed tree will be a tree if restricted to the Dirichlet components. For practical implementation, we further refine the hierarchy by first extending the tree onto the cross-edges before addressing the remaining parts of the wire basket.

\begin{figure}
        \centering
        \begin{tikzpicture}[]
            \node (image) at (0,0) {\includegraphics[width=0.75\linewidth]{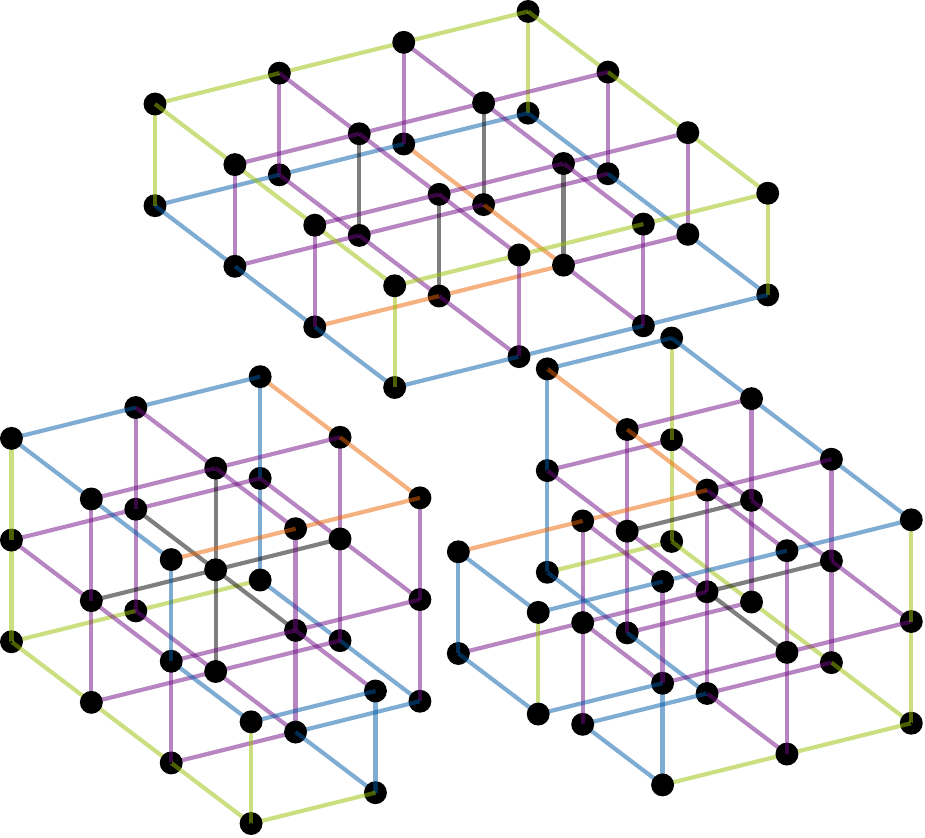}};
            \node at (image.north west) {
                \begin{axis}[%
                    hide axis,
                    xmin=0,
                    xmax=1,
                    ymin=0,
                    ymax=1,
                    legend style={draw=white!15!black,legend cell align=left,{at={(-0.25,-0.1)}},anchor=north west}
                    ]
                    \addlegendimage{ultra thick, weightOne}
                    \addlegendentry{weight 1}
                    \addlegendimage{ultra thick, weightTwo}
                    \addlegendentry{weight 2}
                    \addlegendimage{ultra thick, weightThree}
                    \addlegendentry{weight 3}
                    \addlegendimage{ultra thick, weightFour}
                    \addlegendentry{weight 4}
                    \addlegendimage{ultra thick, weightFive}
                    \addlegendentry{weight 5}
                \end{axis}
            };
        \end{tikzpicture}
        \caption{Exploded view of subdomain graphs corresponding to the decomposition shown in Fig.~\ref{fig:decomp}.}
        \label{fig:graphs}
    \end{figure}

\begin{figure}
        \centering
        \begin{tikzpicture}[]
            \node (image) at (0,0) {\includegraphics[width=0.75\linewidth]{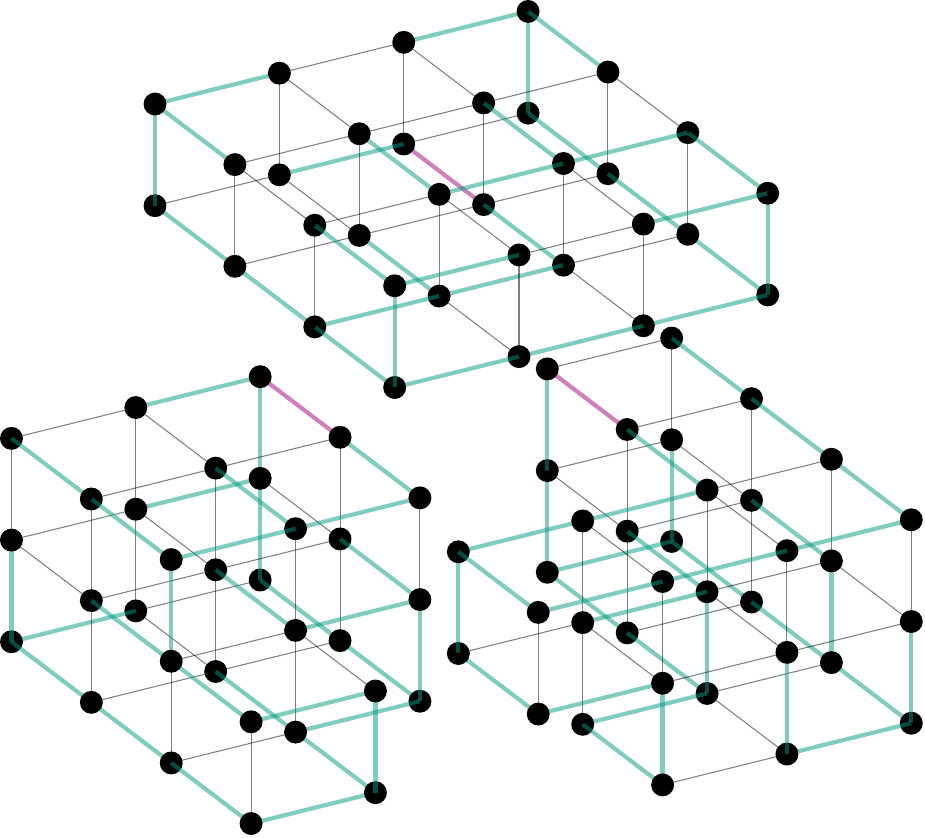}};
            \node at (image.north west) {
                \begin{axis}[%
                    hide axis,
                    xmin=0,
                    xmax=1,
                    ymin=0,
                    ymax=1,
                    legend style={draw=white!15!black,legend cell align=left,{at={(-0.2,-0.45)}},anchor=north west}
                    ]
                    \addlegendimage{ultra thick, treeDOF}
                    \addlegendentry{tree DOFs}
                    \addlegendimage{ultra thick, primalDOF}
                    \addlegendentry{primal DOFs}
                \end{axis}
            };
        \end{tikzpicture}
        \caption{Exploded view of subdomain graphs corresponding to the decomposition shown in Fig.~\ref{fig:decomp}.}
        \label{fig:specialDOFs}
    \end{figure}

    The described hierarchy can be obtained by employing Kruskal's algorithm \cite{Kruskal_1956aa,Cormen_2001aa} with appropriate edge-weights. In Kruskal's algorithm edges with minimal weight are added to the spanning tree, if they do not close a loop with edges which were previously added. In accordance with \cite{Mally_2024ae}, we obtain the weights provided in Fig.~\ref{fig:weights}, which are visualized in Fig.~\ref{fig:graphs} for the exemplary decomposition shown in Fig.~\ref{fig:decomp}.
    \begin{figure}
        \centering
        \fbox{
            \parbox{0.8\linewidth}{
                \begin{itemize}
                    \setlength\itemsep{3mm}
                    \item[\textcolor{TUDa-1b}{\textbf{Weight 1}}] \hspace{1cm} for edges on the intersection of Dirichlet boundary and interfaces
                    \item[\textcolor{TUDa-8b}{\textbf{Weight 2}}] \hspace{1cm} for edges which are shared by more than two subdomains
                    \item[\textcolor{TUDa-4b}{\textbf{Weight 3}}] \hspace{1cm} for remaining edges on the wire basket
                    \item[\textcolor{TUDa-11b}{\textbf{Weight 4}}] \hspace{1cm} for remaining edges on Dirichlet boundary and interfaces
                    \item[\textcolor{black}{\textbf{Weight 5}}] \hspace{1cm} for remaining edges in subdomain interiors
                \end{itemize}
            }
        }
        \caption{Weights to construct consistent spanning tree for application in magnetostatic TI problems.}
        \label{fig:weights}
    \end{figure}
    To resolve additional issues with cross-edges (edges shared by more than two subdomains), we select all cotree edges with weight 2 as primal. This selection, combined with an elimination of tree DOFs, enables both local and global invertibility, i.e., computations exploiting the dual-primal TI approach \eqref{eq:intProb}-\eqref{eq:localProb}. Both tree and primal edges are visualized in Fig.~\ref{fig:specialDOFs} for the decomposition from Fig.~\ref{fig:decomp}.
    In \cite{Mally_2024ae}, all relevant proofs are derived under the assumption that each patch acts as a subdomain. Using multiple patches as subdomains can be interpreted as being equivalent to a slightly modified TI approach from the one presented in \cite{Mally_2024ae}. One difference is that the patch-interface DOFs in the interior of the subdomains are additionally selected as primal DOFs. The other one is that we consider only a subset of the original wire basket of the multipatch approach. Together with the results from \cite{Mally_2024ae}, these aspects ensure solvability.
    Furthermore, it is even implied that the number of primal DOFs of the subdomain decomposition grows linearly with the number of subdomains $N_{\mathrm{sub}}$, but does not depend on the local mesh size $h$ and the degree $p$. Note that different subdomain configurations may lead to a different number of primal DOFs, even if they have the same number of subdomains.

    \section{Numerical Experiments}\label{sec:numExp}
    In the following, we analyze convergence behavior and scalability indicators for the domain configuration depicted in Fig.~\ref{fig:decomp}. To this end, we prescribe boundary conditions and the source term such that the closed-form solution of the strong problem \eqref{eq:strongMS} is given by
    \begin{align*}
        &\Afield_{\mathrm{ana}}(x,y,z)=
        \begin{bmatrix}
            \cos(y)\cos(z)\sin(x) \\
            -2\cos(x)\cos(z)\sin(y) \\
            \cos(x)\cos(y)\sin(z) \\
        \end{bmatrix}\\[1mm]
        &\Bfield_{\mathrm{ana}}(x,y,z)=
        \Curl\Afield_{\mathrm{ana}}=
        \begin{bmatrix}
            -3\cos(x)\sin(y)\sin(z) \\
            0 \\
            3\cos(z)\sin(x)\sin(y) \\
        \end{bmatrix}.
    \end{align*}
    We assume a homogeneous material distribution with $\reluctivity=1$. To quantify the error between the numerical and analytical solutions, we use the $\Hcurl{\Omega}$ semi-norm, equivalent to the $L^2$-error of the magnetic flux density
    \begin{equation}
        \errBfield=\sqrt{\sum_{k=1}^{N_{\mathrm{sub}}}\norm{\Bfield_{\mathrm{ana}}-\Bfield_{\mathrm{num}}}{L^2(\Omega^{(k)})}^2}.
    \end{equation}
    Here, we expect to observe the optimal convergence behavior $\errBfield=\mathcal{O}(h^p)$, as demonstrated in \cite{Mally_2024ae}. Scalability is mainly influenced by the sequential components of the algorithm \cite{Gustafson_1988aa}. In the context of dual-primal TI methods, the major bottlenecks are the factorization of $\coarseMat$ and the iterative scheme (PCG) used to solve \eqref{eq:intProb}. For the first bottleneck the number of primal DOFs $\nCoarse$ is an appropriate indicator. The sequential effort of the PCG approach is determined by the number of iterations $N_{\mathrm{iter}}$, which is intrinsically linked to the condition number $\cond{\mathbf{M}_{\mathrm{D}}^{-1}\interfaceMat}$. For the conditioning, we expect that a bound of the form
    \begin{equation}
        \cond{\mathbf{M}_{\mathrm{D}}^{-1}\interfaceMat}=\frac{\omega_{\mathrm{max}}}{\omega_{\mathrm{min}}}\leq C \left(1+\log_{10}\left(p^2\frac{H}{h}\right)\right)^2=C\left(C_{\mathrm{H}}+\log_{10}\left(\frac{p^2}{h}\right)\right)^2\label{eq:condEstimate}
    \end{equation}
    is satisfied as proposed in \cite{Klawonn_2008aa}. Here, $C$ is a positive constant independent of mesh size $h$, subdomain size $H$ and degree $p$. In this case, we compute the condition number with the largest absolute eigenvalue $\omega_{\mathrm{max}}$ and smallest absolute eigenvalue $\omega_{\mathrm{min}}$. Although the bound of \cite{Klawonn_2008aa} has not been analytically derived for magnetostatic IETI problems, it provides a useful reference for optimal behavior and indicates how our method compares in terms of conditioning. It is often assumed that the number of iterations of the PCG method $N_{\mathrm{iter}}$, to reach a certain tolerance $\eta_{\mathrm{tol}}$, scales proportionally with the square root of the condition number. In other words, $N_{\mathrm{iter}}\sim\sqrt{\cond{\mathbf{M}_{\mathrm{D}}^{-1}\interfaceMat}}$ in our setting, e.g., see \cite[Chapter~10]{Shewchuk_1994aa}. Consequently, we aim to verify that
    \begin{equation}
        N_{\mathrm{iter}} \leq \tilde{C}\left(C_{\mathrm{H}}+\log_{10}\left(\frac{p^2}{h}\right)\right)\label{eq:iterEstimate}
    \end{equation}
    holds.

\begin{figure}
        \centering
        \begin{subfigure}[B]{0.49\linewidth}
            \centering
            \begin{tikzpicture}
                \begin{semilogxaxis}[
                    xmin=3e-2, xmax=2e-1, ymin=0, ymax=2, xlabel={Mesh Size $h$}, ylabel={Size $\nCoarse$}, width=0.95\linewidth, height=5cm, font=\small, ytick={0.5,1,1.5}, yticklabels={0.5,1,1.5}, tudalineplot, xtick={1/24,1/12,1/6}, xticklabels={$\frac{1}{24}$,$\frac{1}{12}$,$\frac{1}{6}$}, legend style = {font=\scriptsize}
                ]
                    \addplot+[mark=square, blue, mark size=2pt] table [
                        col sep=comma,
                        x expr= {1/(\thisrow{divs}*\thisrow{patchs}^(1/3))},
                        y expr = {ifthenelse(and(\thisrow{deg}==1,\thisrow{subs}==3),\thisrow{pri},NaN)},
                    ] {data/decomp_data.csv};
                    \addplot+[mark=diamond, red, mark size=2pt] table [
                        col sep=comma,
                        x expr= {1/(\thisrow{divs}*\thisrow{patchs}^(1/3))},
                        y expr = {ifthenelse(and(\thisrow{deg}==2,\thisrow{subs}==3),\thisrow{pri},NaN)},
                    ] {data/decomp_data.csv};
                    \addplot+[mark=x, cyan, mark size=2pt] table [
                        col sep=comma,
                        x expr= {1/(\thisrow{divs}*\thisrow{patchs}^(1/3))},
                        y expr = {ifthenelse(and(\thisrow{deg}==3,\thisrow{subs}==3),\thisrow{pri},NaN)},
                    ] {data/decomp_data.csv};
                    \legend{$p=1$,$p=2$,$p=3$};
                \end{semilogxaxis}
            \end{tikzpicture}
            \caption{Coarse problem size.}
            \label{fig:exp1_pri}
        \end{subfigure}
        \begin{subfigure}[B]{0.49\linewidth}
            \centering
            \begin{tikzpicture}
                \begin{loglogaxis}[
                    xmin=3e-2, xmax=2e-1, ymin=1e-5, ymax=1e1, xlabel={Mesh Size $h$}, ylabel={Error $\errBfield$}, width=0.95\linewidth, height=5cm, font=\small, ytick={1e-4,1e-2,1e-0}, xtick={1/24,1/12,1/6}, xticklabels={$\frac{1}{24}$,$\frac{1}{12}$,$\frac{1}{6}$}, tudalineplot, legend style = {font=\scriptsize, anchor=south east, at={(0.95,0.05)}}
                ]
                    \addplot+[mark=square, blue, mark size=2pt,forget plot] table [
                        col sep=comma,
                        x expr= {1/(\thisrow{divs}*\thisrow{patchs}^(1/3))},
                        y expr = {ifthenelse(and(\thisrow{deg}==1,\thisrow{subs}==3),\thisrow{err},NaN)},
                    ] {data/decomp_data.csv};
                    \addplot+[mark=diamond, red, mark size=2pt,forget plot] table [
                        col sep=comma,
                        x expr= {1/(\thisrow{divs}*\thisrow{patchs}^(1/3))},
                        y expr = {ifthenelse(and(\thisrow{deg}==2,\thisrow{subs}==3),\thisrow{err},NaN)},
                    ] {data/decomp_data.csv};
                    \addplot+[mark=x, cyan, mark size=2pt,forget plot] table [
                        col sep=comma,
                        x expr= {1/(\thisrow{divs}*\thisrow{patchs}^(1/3))},
                        y expr = {ifthenelse(and(\thisrow{deg}==3,\thisrow{subs}==3),\thisrow{err},NaN)},
                    ] {data/decomp_data.csv};

                    \addplot+[black, dashed, mark=none, samples at={0.04,0.17}] {10^(1.2+log10(x))};
                    \addplot+[black, dashed, mark=none, samples at={0.04,0.17},forget plot] {10^(1.0+2*log10(x))};
                    \addplot+[black, dashed, mark=none, samples at={0.04,0.17},forget plot] {10^(0.6+3*log10(x))};

                    \legend{$Ch^p$};
                \end{loglogaxis}
            \end{tikzpicture}
            \caption{Magnetic flux density error.}
            \label{fig:exp1_err}
        \end{subfigure}
        \\[3mm]
        \begin{subfigure}[B]{0.49\linewidth}
            \centering
            \begin{tikzpicture}
                \begin{semilogxaxis}[
                    xmin=3e-2, xmax=2e-1, ymin=1.5, ymax=5.5, xlabel={Mesh Size $h$}, ylabel={Conditioning $\cond{\mathbf{M}_{\mathrm{D}}^{-1}\interfaceMat}$}, width=0.95\linewidth, height=5cm, font=\small, tudalineplot, xtick={1/24,1/12,1/6}, xticklabels={$\frac{1}{24}$,$\frac{1}{12}$,$\frac{1}{6}$}, legend style = {font=\tiny, anchor=south west, at={(0.05,0.05)}}
                ]
                    \addplot+[mark=square, blue, mark size=2pt,forget plot] table [
                        col sep=comma,
                        x expr= {1/(\thisrow{divs}*\thisrow{patchs}^(1/3))},
                        y expr = {ifthenelse(and(\thisrow{deg}==1,\thisrow{subs}==3),\thisrow{cond},NaN)},
                    ] {data/decomp_data.csv};
                    \addplot+[mark=diamond, red, mark size=2pt,forget plot] table [
                        col sep=comma,
                        x expr= {1/(\thisrow{divs}*\thisrow{patchs}^(1/3))},
                        y expr = {ifthenelse(and(\thisrow{deg}==2,\thisrow{subs}==3),\thisrow{cond},NaN)},
                    ] {data/decomp_data.csv};
                    \addplot+[mark=x, cyan, mark size=2pt,forget plot] table [
                        col sep=comma,
                        x expr= {1/(\thisrow{divs}*\thisrow{patchs}^(1/3))},
                        y expr = {ifthenelse(and(\thisrow{deg}==3,\thisrow{subs}==3),\thisrow{cond},NaN)},
                    ] {data/decomp_data.csv};

                    \addplot+[black, dashed, mark=none, domain=0.04:0.17, samples=101] {0.33*(2.0+log10(1/x))^2};
                    \addplot+[black, dashed, mark=none, domain=0.04:0.17, samples=101] {0.29*(2.0+log10(4/x))^2};
                    \addplot+[black, dashed, mark=none, domain=0.04:0.17, samples=101] {0.27*(2.0+log10(9/x))^2};

                    \legend{$C\left(2.0+\log\left(p^2/h\right)\right)^2$};
                \end{semilogxaxis}
            \end{tikzpicture}
            \caption{Conditioning of precond. interface problem.}
            \label{fig:exp1_cond}
        \end{subfigure}
        \begin{subfigure}[B]{0.49\linewidth}
            \centering
            \begin{tikzpicture}
                \begin{semilogxaxis}[
                    xmin=3e-2, xmax=2e-1, ymin=7, ymax=15, xlabel={Mesh Size $h$}, ylabel={PCG Iter. $N_{\mathrm{iter}}$}, width=0.95\linewidth, height=5cm, font=\small, tudalineplot, xtick={1/24,1/12,1/6}, xticklabels={$\frac{1}{24}$,$\frac{1}{12}$,$\frac{1}{6}$}, legend style = {font=\tiny, anchor=south west, at={(0.05,0.05)}}
                ]
                    \addplot+[mark=square, blue, mark size=2pt,forget plot] table [
                        col sep=comma,
                        x expr= {1/(\thisrow{divs}*\thisrow{patchs}^(1/3))},
                        y expr = {ifthenelse(and(\thisrow{deg}==1,\thisrow{subs}==3),\thisrow{iter},NaN)},
                    ] {data/decomp_data.csv};
                    \addplot+[mark=diamond, red, mark size=2pt,forget plot] table [
                        col sep=comma,
                        x expr= {1/(\thisrow{divs}*\thisrow{patchs}^(1/3))},
                        y expr = {ifthenelse(and(\thisrow{deg}==2,\thisrow{subs}==3),\thisrow{iter},NaN)},
                    ] {data/decomp_data.csv};
                    \addplot+[mark=x, cyan, mark size=2pt,forget plot] table [
                        col sep=comma,
                        x expr= {1/(\thisrow{divs}*\thisrow{patchs}^(1/3))},
                        y expr = {ifthenelse(and(\thisrow{deg}==3,\thisrow{subs}==3),\thisrow{iter},NaN)},
                    ] {data/decomp_data.csv};

                    \addplot+[black, dashed, mark=none, domain=0.04:0.17, samples=101] {3.5*(2.0+log10(1/x))};
                    \addplot+[black, dashed, mark=none, domain=0.04:0.17, samples=101,forget plot] {3.2*(2.0+log10(4/x))};
                    \addplot+[black, dashed, mark=none, domain=0.04:0.17, samples=101,forget plot] {3.2*(2.0+log10(9/x))};

                    \legend{$C\left(2.0+\log\left(p^2/h\right)\right)$};
                \end{semilogxaxis}
            \end{tikzpicture}
            \caption{Number of PCG iterations.}
            \label{fig:exp1_iter}
        \end{subfigure}
        \caption{Scalability and convergence results with respect to mesh size for the domain configuration shown in Fig.~\ref{fig:decomp} with PCG tolerance $\eta_{\mathrm{tol}}=10^{-6}$.}
        \label{fig:exp1}
    \end{figure}
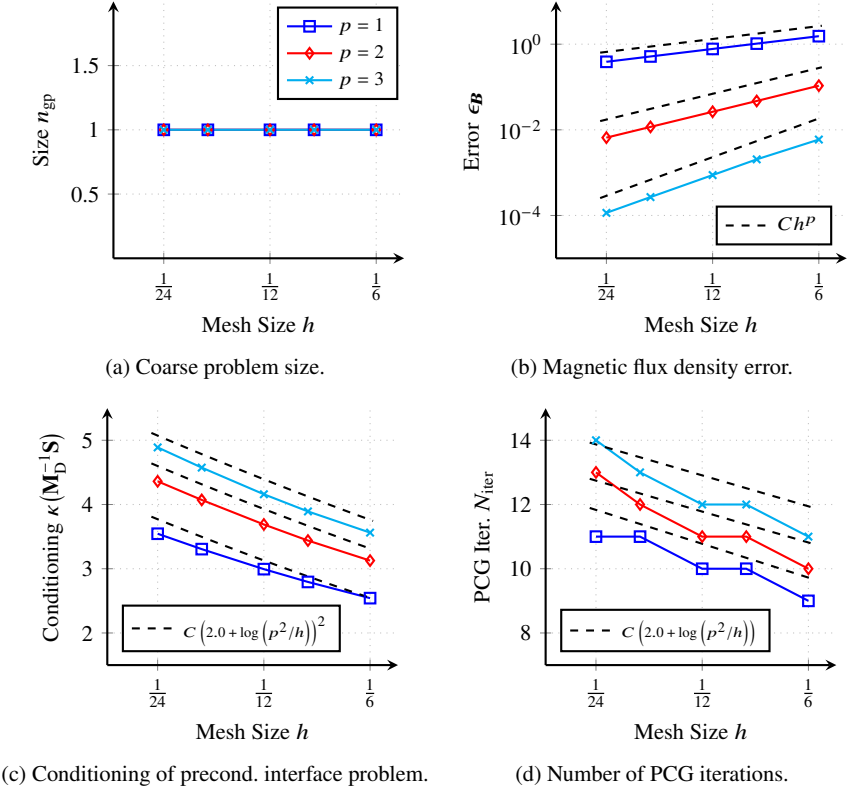

    In the first experiment, we vary the mesh size $h$ and the degree $p$ for the subdomain configuration shown in Fig.~\ref{fig:decomp} consisting of $N_{\mathrm{sub}}=3$ subdomains and 27 patches. The results for this experiment are presented in Fig.~\ref{fig:exp1}. We observe in Fig.~\ref{fig:exp1_pri} that the coarse problem size $\nCoarse$ is independent of $h$ and $p$, as expected. In Fig.~\ref{fig:exp1_err}, we see optimal convergence behavior of the error $\errBfield$. Fig.~\ref{fig:exp1_cond} confirms that the condition number satisfies the bound \eqref{eq:condEstimate}, which is also consistent with the optimal TI estimates for $p=1$ in the scalar case, see e.g., \cite{Farhat_2001aa}.
    Finally, Fig.~\ref{fig:exp1_iter} shows that the required amount of PCG iterations $N_{\mathrm{iter}}$ are in good agreement with the estimate \eqref{eq:iterEstimate}.

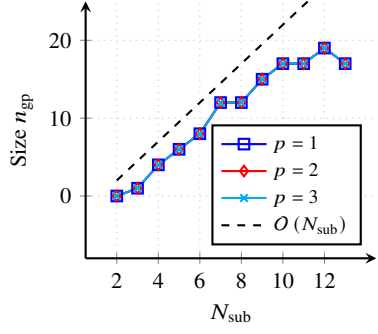
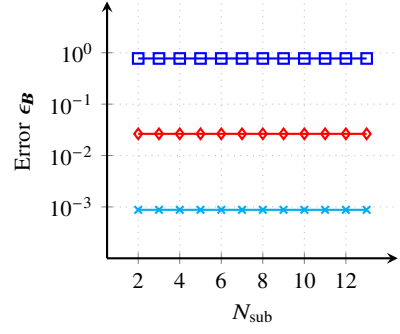
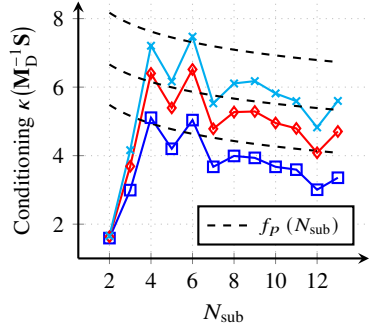
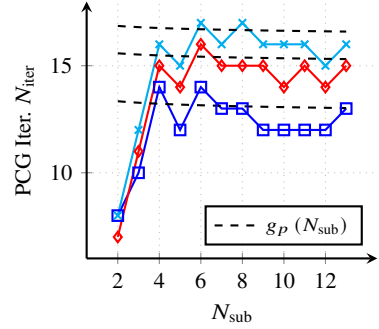
\begin{figure}
        \centering
        \begin{subfigure}[B]{0.49\linewidth}
            \centering
            \begin{tikzpicture}
                \begin{axis}[
                    xmin=0.5, xmax=14.5, ymin=-8, ymax=25, xlabel={$N_{\mathrm{sub}}$}, ylabel={Size $\nCoarse$}, width=0.95\linewidth, height=5cm, font=\small, ytick={0,10,20}, yticklabels={0,10,20}, tudalineplot, xtick={2,4,6,8,10,12}, xticklabels={2,4,6,8,10,12}, legend style = {font=\scriptsize, anchor=south east, at={(0.95,0.05)}}
                ]
                    \addplot+[mark=square, blue, mark size=2pt] table [
                        col sep=comma,
                        x expr= {\thisrow{subs}},
                        y expr = {ifthenelse(and(\thisrow{deg}==1,\thisrow{divs}==4),\thisrow{pri},NaN)},
                    ] {data/decomp_data.csv};
                    \addplot+[mark=diamond, red, mark size=2pt] table [
                        col sep=comma,
                        x expr= {\thisrow{subs}},
                        y expr = {ifthenelse(and(\thisrow{deg}==2,\thisrow{divs}==4),\thisrow{pri},NaN)},
                    ] {data/decomp_data.csv};
                    \addplot+[mark=x, cyan, mark size=2pt] table [
                        col sep=comma,
                        x expr= {\thisrow{subs}},
                        y expr = {ifthenelse(and(\thisrow{deg}==3,\thisrow{divs}==4),\thisrow{pri},NaN)},
                    ] {data/decomp_data.csv};
                    \addplot+[black, dashed, mark=none, samples at={2,13}] {2.5*x-3};
                    \legend{$p=1$,$p=2$,$p=3$,$\mathcal{O}\left(N_{\mathrm{sub}}\right)$};
                \end{axis}
            \end{tikzpicture}
            \caption{Coarse problem size.}
            \label{fig:exp2_pri}
        \end{subfigure}
        \begin{subfigure}[B]{0.49\linewidth}
            \centering
            \begin{tikzpicture}
                \begin{semilogyaxis}[
                    xmin=0.5, xmax=14.5, ymin=1e-4, ymax=1e1, xlabel={$N_{\mathrm{sub}}$}, ylabel={Error $\errBfield$}, width=0.95\linewidth, height=5cm, font=\small, ytick={1e-3,1e-2,1e-1,1e-0}, xtick={2,4,6,8,10,12}, xticklabels={2,4,6,8,10,12}, yminorticks=false, tudalineplot
                ]
                    \addplot+[mark=square, blue, mark size=2pt,forget plot] table [
                        col sep=comma,
                        x expr= {\thisrow{subs}},
                        y expr = {ifthenelse(and(\thisrow{deg}==1,\thisrow{divs}==4),\thisrow{err},NaN)},
                    ] {data/decomp_data.csv};
                    \addplot+[mark=diamond, red, mark size=2pt,forget plot] table [
                        col sep=comma,
                        x expr= {\thisrow{subs}},
                        y expr = {ifthenelse(and(\thisrow{deg}==2,\thisrow{divs}==4),\thisrow{err},NaN)},
                    ] {data/decomp_data.csv};
                    \addplot+[mark=x, cyan, mark size=2pt,forget plot] table [
                        col sep=comma,
                        x expr= {\thisrow{subs}},
                        y expr = {ifthenelse(and(\thisrow{deg}==3,\thisrow{divs}==4),\thisrow{err},NaN)},
                    ] {data/decomp_data.csv};
                \end{semilogyaxis}
            \end{tikzpicture}
            \caption{Magnetic flux density error.}
            \label{fig:exp2_err}
        \end{subfigure}
        \\[3mm]
        \begin{subfigure}[B]{0.49\linewidth}
            \centering
            \begin{tikzpicture}
                \begin{axis}[
                    xmin=0.5, xmax=14.5, ymin=1, ymax=8.5, xlabel={$N_{\mathrm{sub}}$}, ylabel={Conditioning $\cond{\mathbf{M}_{\mathrm{D}}^{-1}\interfaceMat}$}, width=0.95\linewidth, height=5cm, font=\small, tudalineplot, xtick={2,4,6,8,10,12}, xticklabels={2,4,6,8,10,12}, legend style = {font=\scriptsize, anchor=south east, at={(0.95,0.05)}}
                ]
                    \addplot+[mark=square, blue, mark size=2pt,forget plot] table [
                        col sep=comma,
                        x expr= {\thisrow{subs}},
                        y expr = {ifthenelse(and(\thisrow{deg}==1,\thisrow{divs}==4),\thisrow{cond},NaN)},
                    ] {data/decomp_data.csv};
                    \addplot+[mark=diamond, red, mark size=2pt,forget plot] table [
                        col sep=comma,
                        x expr= {\thisrow{subs}},
                        y expr = {ifthenelse(and(\thisrow{deg}==2,\thisrow{divs}==4),\thisrow{cond},NaN)},
                    ] {data/decomp_data.csv};
                    \addplot+[mark=x, cyan, mark size=2pt,forget plot] table [
                        col sep=comma,
                        x expr= {\thisrow{subs}},
                        y expr = {ifthenelse(and(\thisrow{deg}==3,\thisrow{divs}==4),\thisrow{cond},NaN)},
                    ] {data/decomp_data.csv};

                    \addplot+[black, dashed, mark=none, domain=2:13, samples=101] {1.4*(1-log10(x)/3+log10(12))^2};
                    \addplot+[black, dashed, mark=none, domain=2:13, samples=101, forget plot] {(1-log10(x)/3+log10(12*4))^2};
                    \addplot+[black, dashed, mark=none, domain=2:13, samples=101, forget plot] {0.95*(1-log10(x)/3+log10(12*9))^2};

                    \legend{$f_p\left(N_{\mathrm{sub}}\right)$};

                \end{axis}
            \end{tikzpicture}
            \caption{Conditioning of precond. interface problem.}
            \label{fig:exp2_cond}
        \end{subfigure}
        \begin{subfigure}[B]{0.49\linewidth}
            \centering
            \begin{tikzpicture}
                \begin{axis}[
                    xmin=0.5, xmax=14.5, ymin=6, ymax=18, xlabel={$N_{\mathrm{sub}}$}, ylabel={PCG Iter. $N_{\mathrm{iter}}$}, width=0.95\linewidth, height=5cm, font=\small, tudalineplot, xtick={2,4,6,8,10,12}, xticklabels={2,4,6,8,10,12}, legend style = {font=\scriptsize, anchor=south east, at={(0.95,0.05)}}
                ]
                    \addplot+[mark=square, blue, mark size=2pt,forget plot] table [
                        col sep=comma,
                        x expr= {\thisrow{subs}},
                        y expr = {ifthenelse(and(\thisrow{deg}==1,\thisrow{divs}==4),\thisrow{iter},NaN)},
                    ] {data/decomp_data.csv};
                    \addplot+[mark=diamond, red, mark size=2pt,forget plot] table [
                        col sep=comma,
                        x expr= {\thisrow{subs}},
                        y expr = {ifthenelse(and(\thisrow{deg}==2,\thisrow{divs}==4),\thisrow{iter},NaN)},
                    ] {data/decomp_data.csv};
                    \addplot+[mark=x, cyan, mark size=2pt,forget plot] table [
                        col sep=comma,
                        x expr= {\thisrow{subs}},
                        y expr = {ifthenelse(and(\thisrow{deg}==3,\thisrow{divs}==4),\thisrow{iter},NaN)},
                    ] {data/decomp_data.csv};

                    \addplot+[black, dashed, mark=none, domain=2:13, samples=101] {sqrt(1.4*(1-log10(x)/3+log10(12))^2)+11};
                    \addplot+[black, dashed, mark=none, domain=2:13, samples=101, forget plot] {sqrt((1-log10(x)/3+log10(12*4))^2)+13};
                    \addplot+[black, dashed, mark=none, domain=2:13, samples=101, forget plot] {sqrt(0.95*(1-log10(x)/3+log10(12*9))^2)+14};

                    \legend{$g_p\left(N_{\mathrm{sub}}\right)$};
                \end{axis}
            \end{tikzpicture}
            \caption{Number of PCG iterations.}
            \label{fig:exp2_iter}
        \end{subfigure}
        \caption{Scalability and convergence results with respect to number of subdomains for automatically generated domain configurations for $h=\frac{1}{12}$ with PCG tolerance $\eta_{\mathrm{tol}}=10^{-6}$.}
        \label{fig:exp2}
    \end{figure}

    In the second experiment, we vary the number of subdomains $N_{\mathrm{sub}}\in\{2,\ldots,12\}$ and the degree $p$ while keeping the local mesh size constant at $h = \frac{1}{12}$. Here, we use automatically generated subdomain configurations for the multipatch structure with 27 patches. The results of this experiment are summarized in Fig.~\ref{fig:exp2}. The coarse problem size $\nCoarse$ can be bounded from above with a linear function of $N_{\mathrm{sub}}$ as visualized in Fig.~\ref{fig:exp2_pri}. This is consistent with the theoretical results in \cite{Mally_2024ae}. We expect the error $\errBfield$, to remain constant for fixed $p$, as the underlying multipatch problem remains the unchanged when $h$ is constant. This expectation is confirmed by the numerical results in Fig.~\ref{fig:exp2_err}.
    Finally, the condition number and number of PCG iterations agree with the theoretical predictions \eqref{eq:condEstimate} and \eqref{eq:iterEstimate}. This is illustrated in Figs.~\ref{fig:exp2_cond} and \ref{fig:exp2_iter}, where we show
    \begin{equation*}
        f_p\left(N_{\mathrm{sub}}\right)=a_p\left(1-\frac{\log_{10}\left(N_{\mathrm{sub}}\right)}{3}+\log_{10}\left(12p^2\right)\right)^2
    \end{equation*}
    and
    \begin{equation*}
        g_p\left(N_{\mathrm{sub}}\right)=\sqrt{f_p\left(N_{\mathrm{sub}}\right)}+b_p.
    \end{equation*}
    Note that we implicitly assumed $H\leq C N_{\mathrm{sub}}^{-\frac{1}{3}}$ for some positive constant $C\in\mathbb{R}_{>0}$ and that $a_p,b_p\in\mathbb{R}_{>0}$ are constants chosen appropriately for every $p\in\{1,2,3\}$.

    \section{Conclusion}
    In this work, we extended the dual-primal IETI approach for 3D magnetostatics from \cite{Mally_2024ae} to subdomains consisting of multiple IGA-patches. All results concerning solvability and scalability from \cite{Mally_2024ae} can be translated to this generalized setting, as the global DD problem can be considered as a special case of the DD formulation where each patch is considered to be an individual subdomain. These theoretical considerations are validated using an academic problem by examining convergence properties and typical scalability indicators.

    Overall, we observe that the tree-cotree based IETI approach shows optimal behavior in the context of the tearing and interconnecting framework. This paves the way for automatic load-balancing using graph partitioning approaches. However, a notable challenge remains. Addressing the treatment of subdomains that are not simply-connected or excluding them from the beginning by enforcing the graph partitioning scheme to provide simply-connected subdomains. To the best of our knowledge, both aspects have not yet been researched, making them promising topics for future work.

    \begin{acknowledgement}
        The work is supported by the joint DFG/FWF Collaborative Research Centre CREATOR (DFG: Project-ID 492661287/TRR 361; FWF: 10.55776/F90) at TU Darmstadt, TU Graz and JKU Linz. We acknowledge the funding of The ``Ernst Ludwig Mobility Grant'' of the Association of Friends of Technical University of Darmstadt e.V.
    \end{acknowledgement}

    \bibliographystyle{abbrvnat}
    \renewcommand{\bibnumfmt}[1]{#1.}
    \bibliography{bibtex}

\begin{thebibliography}{33}
\providecommand{\natexlab}[1]{#1}
\providecommand{\url}[1]{\texttt{#1}}
\expandafter\ifx\csname urlstyle\endcsname\relax
  \providecommand{\doi}[1]{doi: #1}\else
  \providecommand{\doi}{doi: \begingroup \urlstyle{rm}\Url}\fi

\bibitem[Albanese and Rubinacci(1988)]{Albanese_1988aa}
R.~Albanese and G.~Rubinacci.
\newblock Integral formulation for {3D} eddy-current computation using edge
  elements.
\newblock \emph{{IEE} Proc. Sci. Meas. Tech.}, 135\penalty0 (7):\penalty0
  457--462, 09 1988.
\newblock ISSN 1350-2344.
\newblock \doi{10.1049/ip-a-1:19880072}.

\bibitem[Arnold(2018)]{Arnold_2018aa}
D.~N. Arnold.
\newblock \emph{Finite Element Exterior Calculus}.
\newblock SIAM, 2018.
\newblock ISBN 978-1-61197-553-6.
\newblock \doi{10.1137/1.9781611975543}.

\bibitem[Bosy et~al.(2020)Bosy, Montardini, Sangalli, and Tani]{Bosy_2020aa}
M.~Bosy, M.~Montardini, G.~Sangalli, and M.~Tani.
\newblock A domain decomposition method for isogeometric multi-patch problems
  with inexact local solvers.
\newblock \emph{Comput. Math. Appl.}, 80\penalty0 (11):\penalty0 2604--2621, 12
  2020.
\newblock ISSN 0898-1221.
\newblock \doi{10.1016/j.camwa.2020.08.024}.

\bibitem[Bouclier and Hirschler(2022)]{Bouclier_2022aa}
R.~Bouclier and T.~Hirschler.
\newblock \emph{{IGA}: Non-conforming coupling and shape optimization of
  complex multipatch structures}, volume~1.
\newblock ISTE Ltd and John Wiley \& Sons, 10 2022.

\bibitem[Buffa et~al.(2010)Buffa, Sangalli, and Vázquez]{Buffa_2010aa}
A.~Buffa, G.~Sangalli, and R.~Vázquez.
\newblock Isogeometric analysis in electromagnetics: {B}-splines approximation.
\newblock \emph{Comput. Meth. Appl. Mech. Eng.}, 199:\penalty0 1143--1152,
  2010.
\newblock ISSN 0045-7825.
\newblock \doi{10.1016/j.cma.2009.12.002}.

\bibitem[Buffa et~al.(2015)Buffa, Vázquez, Sangalli, and Beirão~da
  Veiga]{Buffa_2015aa}
A.~Buffa, R.~H. Vázquez, G.~Sangalli, and L.~Beirão~da Veiga.
\newblock Approximation estimates for isogeometric spaces in multipatch
  geometries.
\newblock \emph{Numer. Meth. Part. Differ. Equat.}, 31\penalty0 (2):\penalty0
  422--438, 2015.
\newblock \doi{10.1002/num.21943}.

\bibitem[Buffa et~al.(2019)Buffa, Dölz, Kurz, Schöps, Vázquez, and
  Wolf]{Buffa_2019ac}
A.~Buffa, J.~Dölz, S.~Kurz, S.~Schöps, R.~Vázquez, and F.~Wolf.
\newblock Multipatch approximation of the de {Rham} sequence and its traces in
  isogeometric analysis.
\newblock \emph{Numer. Math.}, 144\penalty0 (1):\penalty0 201--236, 06 2019.
\newblock ISSN 0029-599X.
\newblock \doi{10.1007/s00211-019-01079-x}.
\newblock arxiv:1806.01062.

\bibitem[Cormen et~al.(2001)Cormen, Leiserson, and Rivest]{Cormen_2001aa}
T.~H. Cormen, C.~E. Leiserson, and R.~L. Rivest.
\newblock \emph{Introduction To Algorithms}.
\newblock MIT Press, 2001.
\newblock ISBN 978-0262032933.

\bibitem[Dostál et~al.(2006)Dostál, Horák, and Kučera]{Dostal_2006aa}
Z.~Dostál, D.~Horák, and R.~Kučera.
\newblock Total {FETI}-an easier implementable variant of the {FETI} method for
  numerical solution of elliptic {PDE}.
\newblock \emph{Comm. Numer. Meth. Eng.}, 22\penalty0 (12):\penalty0
  1155--1162, 06 2006.
\newblock \doi{10.1002/cnm.881}.

\bibitem[Dular et~al.(1995)Dular, Nicolet, Genon, and Legros]{Dular_1995aa}
P.~Dular, A.~Nicolet, A.~Genon, and W.~Legros.
\newblock A discrete sequence associated with mixed finite elements and its
  gauge condition for vector potentials.
\newblock \emph{{IEEE} Trans. Magn.}, 31\penalty0 (3):\penalty0 1356--1359, 05
  1995.
\newblock ISSN 0018-9464.
\newblock \doi{10.1109/20.376278}.

\bibitem[Farhat and Roux(1991)]{Farhat_1991aa}
C.~Farhat and F.-X. Roux.
\newblock A method of finite element tearing and interconnecting and its
  parallel solution algorithm.
\newblock \emph{Int. J. Numer. Meth. Eng.}, 32\penalty0 (6):\penalty0
  1205--1227, 10 1991.
\newblock ISSN 0029-5981.
\newblock \doi{10.1002/nme.1620320604}.

\bibitem[Farhat et~al.(2000)Farhat, Lesoinne, and Pierson]{Farhat_2000aa}
C.~Farhat, M.~Lesoinne, and K.~Pierson.
\newblock A scalable dual-primal domain decomposition method.
\newblock \emph{Numer. Lin. Algebra. Appl.}, 7\penalty0 (7-8):\penalty0
  687--714, 2000.
\newblock \doi{10.1002/1099-1506(200010/12)7:7/8<687::AID-NLA219>3.0.CO;2-S}.

\bibitem[Farhat et~al.(2001)Farhat, Lesoinne, LeTallec, Pierson, and
  Rixen]{Farhat_2001aa}
C.~Farhat, M.~Lesoinne, P.~LeTallec, K.~Pierson, and D.~Rixen.
\newblock {FETI-DP}: a dual–primal unified {FETI} method—part i: A faster
  alternative to the two-level {FETI} method.
\newblock \emph{Int. J. Numer. Meth. Eng.}, 50\penalty0 (7):\penalty0
  1523--1544, 2001.
\newblock ISSN 0029-5981.
\newblock \doi{10.1002/nme.76}.

\bibitem[Ghenai et~al.(2024)Ghenai, Perrussel, Chadebec, Vi, Guichon, Meunier,
  and Siau]{Ghenai_2024aa}
M.~I. Ghenai, R.~Perrussel, O.~Chadebec, F.~Vi, J.-M. Guichon, G.~Meunier, and
  J.~Siau.
\newblock Domain decomposition for {3-D} nonlinear magnetostatic problems:
  {Newton–Krylov–Schur} versus {Schur–Newton–Krylov} methods.
\newblock \emph{{IEEE} Trans. Magn.}, 60\penalty0 (3):\penalty0 1--4, 2024.
\newblock ISSN 0018-9464.
\newblock \doi{10.1109/TMAG.2023.3299989}.

\bibitem[Gustafson(1988)]{Gustafson_1988aa}
J.~L. Gustafson.
\newblock Reevaluating {Amdahl's} law.
\newblock \emph{{CACM}}, 31\penalty0 (5):\penalty0 532--533, 1988.
\newblock ISSN 0001-0782.
\newblock \doi{10.1145/42411.42415}.

\bibitem[Jackson(1998)]{Jackson_1998aa}
J.~D. Jackson.
\newblock \emph{Classical Electrodynamics}.
\newblock Wiley {\&} Sons, New York, NY, USA, 3rd edition, 1998.
\newblock ISBN 978-0-471-30932-1.
\newblock \doi{10.1017/CBO9780511760396}.

\bibitem[Kapidani et~al.(2022)Kapidani, Merkel, Schöps, and
  Vázquez]{Kapidani_2022aa}
B.~Kapidani, M.~Merkel, S.~Schöps, and R.~Vázquez.
\newblock Tree-cotree decomposition of isogeometric mortared spaces in
  {H}(curl) on multi-patch domains.
\newblock \emph{Comput. Meth. Appl. Mech. Eng.}, 395:\penalty0 114949, 05 2022.
\newblock ISSN 0045-7825.
\newblock \doi{10.1016/j.cma.2022.114949}.
\newblock arxiv:2110.15860.

\bibitem[Karypis and Kumar(1997)]{Karypis_1997aa}
G.~Karypis and V.~Kumar.
\newblock {METIS:} a software package for partitioning unstructured graphs,
  partitioning meshes, and computing fill-reducing orderings of sparse
  matrices.
\newblock Technical report, University Digital Conservancy, Computer Science \&
  Engineering {(CS\&E)} Technical Reports, Minnesota, 1997.
\newblock URL \url{https://hdl.handle.net/11299/215346}.

\bibitem[Klawonn et~al.(2002)Klawonn, Widlund, and Dryja]{Klawonn_2002aa}
A.~Klawonn, O.~B. Widlund, and M.~Dryja.
\newblock Dual-primal {FETI} methods for three-dimensional elliptic problems
  with heterogeneous coefficients.
\newblock \emph{{SIAM} J. Numer. Anal.}, 40\penalty0 (1):\penalty0 159--179,
  2002.
\newblock \doi{10.1137/S0036142901388081}.

\bibitem[Klawonn et~al.(2008)Klawonn, Pavarino, and Rheinbach]{Klawonn_2008aa}
A.~Klawonn, L.~F. Pavarino, and O.~Rheinbach.
\newblock Spectral element {FETI-DP} and {BDDC} preconditioners with
  multi-element subdomains.
\newblock \emph{Comput. Meth. Appl. Mech. Eng.}, 198\penalty0 (3):\penalty0
  511--523, 2008.
\newblock ISSN 0045-7825.
\newblock \doi{10.1016/j.cma.2008.08.017}.

\bibitem[Kleiss et~al.(2012)Kleiss, Pechstein, Jüttler, and
  Tomar]{Kleiss_2012ab}
S.~K. Kleiss, C.~Pechstein, B.~Jüttler, and S.~Tomar.
\newblock {IETI} -- isogeometric tearing and interconnecting.
\newblock \emph{Comput. Meth. Appl. Mech. Eng.}, 247:\penalty0 201--215, 11
  2012.
\newblock ISSN 0045-7825.
\newblock \doi{10.1016/j.cma.2012.08.007}.

\bibitem[Kruskal(1956)]{Kruskal_1956aa}
J.~B. Kruskal.
\newblock On the shortest spanning subtree of a graph and the traveling
  salesman problem.
\newblock \emph{{PROC}}, 7\penalty0 (1):\penalty0 48--50, 1956.
\newblock ISSN 1088-6826.
\newblock \doi{10.2307/2033241}.

\bibitem[Mally et~al.(2024)Mally, Kapidani, Merkel, and Schöps]{Mally_2024ae}
M.~Mally, B.~Kapidani, M.~Merkel, and S.~Schöps.
\newblock Tree-cotree-based tearing and interconnecting for {3D}
  magnetostatics: A dual-primal approach.
\newblock Preprint arxiv:2407.21707, Cornell University, 2024.
\newblock submitted.

\bibitem[Manges and Cendes(1995)]{Manges_1995aa}
J.~B. Manges and Z.~J. Cendes.
\newblock A generalized tree-cotree gauge for magnetic field computation.
\newblock \emph{{IEEE} Trans. Magn.}, 31\penalty0 (3):\penalty0 1342--1347,
  1995.
\newblock ISSN 0018-9464.
\newblock \doi{10.1109/20.376275}.

\bibitem[Marcsa and Kuczmann(2013)]{Marcsa_2013aa}
D.~Marcsa and M.~Kuczmann.
\newblock Finite element tearing and interconnecting method and its algorithms
  for parallel solution of magnetic field problems.
\newblock \emph{Electrical, Control and Communication Engineering}, 3\penalty0
  (1):\penalty0 25--30, 08 2013.
\newblock \doi{10.2478/ecce-2013-0011}.

\bibitem[Merkel et~al.(2022)Merkel, Kapidani, Schöps, and
  Vázquez]{Merkel_2022ab}
M.~Merkel, B.~Kapidani, S.~Schöps, and R.~Vázquez.
\newblock Torque computation with the isogeometric mortar method for the
  simulation of electric machines.
\newblock \emph{{IEEE} Trans. Magn.}, 58\penalty0 (9), 06 2022.
\newblock ISSN 0018-9464.
\newblock \doi{10.1109/TMAG.2022.3186247}.
\newblock arxiv:2202.05771.

\bibitem[Monk(2003)]{Monk_2003aa}
P.~Monk.
\newblock \emph{Finite Element Methods for {Maxwell}'s Equations}.
\newblock Oxford University Press, Oxford, 2003.

\bibitem[Of and Steinbach(2009)]{Of_2009ab}
G.~Of and O.~Steinbach.
\newblock The all-floating boundary element tearing and interconnecting method.
\newblock \emph{J. Numer. Math.}, 17\penalty0 (4), 01 2009.
\newblock \doi{10.1515/jnum.2009.014}.

\bibitem[Schneckenleitner and Takacs(2022)]{Schneckenleitner_2022aa}
R.~Schneckenleitner and S.~Takacs.
\newblock {IETI-DP} methods for discontinuous {Galerkin} multi-patch
  isogeometric analysis with {T}-junctions.
\newblock \emph{Comput. Meth. Appl. Mech. Eng.}, 393:\penalty0 114694, 2022.
\newblock ISSN 0045-7825.
\newblock \doi{10.1016/j.cma.2022.114694}.

\bibitem[Shewchuk(1994)]{Shewchuk_1994aa}
J.~R. Shewchuk.
\newblock An introduction to the conjugate gradient method without the
  agonizing pain.
\newblock Technical report, Carnegie Mellon University, Pittsburgh, PA, USA,
  1994.

\bibitem[Toselli(2006)]{Toselli_2006aa}
A.~Toselli.
\newblock Dual-primal {FETI} algorithms for edge finite-element approximations
  in {3D}.
\newblock \emph{{IMA} J. Numer. Anal.}, 26\penalty0 (1):\penalty0 96--130, 01
  2006.
\newblock ISSN 0272-4979.
\newblock \doi{10.1093/imanum/dri023}.

\bibitem[Vázquez(2016)]{Vazquez_2016aa}
R.~Vázquez.
\newblock A new design for the implementation of isogeometric analysis in
  {Octave} and {Matlab}: {GeoPDEs} 3.0.
\newblock \emph{Comput. Math. Appl.}, 72\penalty0 (3):\penalty0 523--554, 08
  2016.
\newblock ISSN 0898-1221.
\newblock \doi{10.1016/j.camwa.2016.05.010}.

\bibitem[Yao et~al.(2012)Yao, Jin, and Krein]{Yao_2012aa}
W.~Yao, J.-M. Jin, and P.~T. Krein.
\newblock Analysis of electromechanical problems using the dual-primal finite
  element tearing and interconnecting method.
\newblock In \emph{2012 {IEEE} Power and Energy Conference at Illinois}, pages
  1--5, 2012.
\newblock \doi{10.1109/PECI.2012.6184595}.

\end{thebibliography}
\end{document}